# Multifunctionality in Janus $CrMCN_4$ (M = Si/Ge) Monolayers: Valleytronic Physics, Piezoelectric Response, and Photocatalytic Potential

Jubair Hossan Abir[1], M.R.U.S. Shapon[1], S.S.B. Pallab[2], Tanvir Khan[1],
Raihana Shams Islam[1*], Saleh Hasan Naqib[1*]

[1]Department of Physics, University of Rajshahi, Rajshahi 6205
[2]Department of Electrical and Electronic Engineering, University of Rajshahi, Rajshahi 6205
*Corresponding authors; Emails: **jsr_eda1@ru.ac.bd** (Saleh Hasan Naqib), **rsislam@ru.ac.bd** (Raihana Shams Islam)

## Abstract

Two-dimensional Janus semiconductors integrating spin-valley coupling, piezoelectricity, and tunable optical responses offer a platform for multifunctional nanodevices. Here, first-principles calculations reveal complementary composition and strain effects in $CrSiCN_4$ and $CrGeCN_4$ monolayers. Both are found to benonmagnetic direct-gap semiconductors, with gaps of 1.23 and 1.09 eV using the Perdew-Burke-Ernzerhof functional including spin-orbit coupling (PBE+SOC), respectively. HSE06 hybrid-functional calculations retain the direct-gap character, yielding gaps of 1.46 eV for $CrSiCN_4$ and 1.19 eV for $CrGeCN_4$. Cr-*d*/N-*p* hybridization governs the band edges, with valence-band spin splitting of 53.3 and 52.4 meV and conduction-band splitting about 1.6 meV. Opposite out-of-plane spin character and Berry curvature emerge at K and K′. Despite similar valley-edge spin splitting, $CrGeCN_4$ exhibits Berry-curvature hot spots of order $10^3$ $bohr^2$, suggesting enhanced multiband contributions. Biaxial strain from −4% to +4% reduces the respective gaps from 1.59 to 0.86 eV and 1.44 to 0.63 eV, drives direct-to-indirect transitions, and redshifts absorption while preserving valley spin contrast. $CrGeCN_4$ exhibits larger in-plane piezoelectric response ($d_{11}$ = 5.084 pm $V^{-1}$), whereas $CrSiCN_4$ exhibits larger out-of-plane magnitude ($|d_{31}|$ = 0.082 pm $V^{-1}$). Band-edge alignment satisfies the oxygen evolution reaction (OER) requirement but provides insufficient driving force for the hydrogen evolution reaction (HER), motivating further strain-induced band-edge modulation toward overall water splitting. These results establish composition and strain as complementary controls for tailoring Cr-based Janus monolayers for valleytronic, optoelectronic, electromechanical, and photocatalytic applications.

## 1 Introduction

In two-dimensional (2D) semiconductors, the valley degree of freedom complements charge and spin as an additional carrier label that can be used for information processing [1]. Their strong in-plane bonding and small thickness also permit substantial elastic deformation, making strain a practical route for modifying electronic and optical properties [2,3]. For these applications, the orbital, spin, and momentum character of the band-edge states can be as important as the magnitude of the band gap.

Valleytronics uses inequivalent extrema in momentum space as an information degree of freedom. In hexagonal monolayers, K and K′ form a time-reversal-related valley pair. When inversion symmetry is broken, the two valleys can carry opposite Berry curvature and orbital magnetic moment; spin-orbit coupling (SOC) can couple spin and valley and generate valley-contrasting spin splitting and polarization dependent optical selection rules [4]. In monolayer $MoS_2$, valley-selective circular dichroism follows from symmetry-controlled optical selection rules [5], optical helicity can create a valley population imbalance [6], and intrinsic valley Hall transport is observed in non-centrosymmetric odd-layer $MoS_2$ but suppressed when inversion symmetry is restored [7]. A microscopic description of a candidate valley material therefore requires its symmetry, orbital character, SOC response, and Berry curvature to be considered together.

Mechanical strain perturbs these quantities by changing lattice geometry, orbital overlap, and the relative energies of competing band extrema. In 2D semiconductors, tensile or compressive strain can shift band edges, change direct/indirect gap ordering, alter carrier transport, and move optical transition energies [2,3]. Relativistic calculations for monolayer transition-metal dichalcogenides further show that biaxial strain can reorganize orbital contributions and SOC-induced spin splittings [8]. Recent studies also demonstrate strain control of valley-related responses in magnetic or alloyed systems: V-alloyed $MoTe_2$ shows strain and electric-field dependent valley polarization [9], ferrovalley $NbX_2$/$TaX_2$ monolayers exhibit strain-tunable valley polarization together with Berry-curvature derived anomalous Hall response [10], and $VSe_2$/VN shows strain-dependent spin filtering and valley splitting alongside an anomalous Hall response [11]. These examples motivate treating strain as a controlled perturbation of the orbital-SOC valley structure rather than only as a band-gap tuning parameter.

The chemical-vapor-deposition synthesis of septuple-layer $MoSi_2N_4$ and $WSi_2N_4$ established a two-dimensional structural motif that does not depend on exfoliation from a layered bulk parent [12]. The architecture can be described as an $MZ_2$-like core embedded between A-Z sublayers and has been generalized to a broad $MA_2Z_4$ family with diverse compositions and electronic phases [13]. Subsequent surveys document electronic, magnetic, thermal-transport, and superconducting behavior across this family, with strain and elemental substitution providing substantial tunability [14]. Chemical substitution and Janus construction can lower the symmetry and create chemically inequivalent out-of-plane environments within the same septuple-layer framework [15]. This combination of compositional flexibility and symmetry control makes $MA_2Z_4$ derivatives well suited for examining how orbital character and electrostatic asymmetry influence SOC and valley physics.

Valley-related phenomena are already established in several $MA_2Z_4$ systems. $MoSi_2N_4$, $WSi_2N_4$, and $MoSi_2As_4$ exhibit spin-valley coupling, valley-contrasting Berry curvature, and circularly selective optical response [16]. Calculations for $MoSi_2X_4$ (X = N, P, As) show that strain can tune the band structure and SOC-driven valley features [17], while studies of $WSi_2N_4$-derived monolayers connect structural symmetry and orbital character with distinct spin-orbit fields and valley responses [18]. Janus derivatives add an out-of-plane asymmetric environment: $MSiGeN_4$ (M = Mo, W) has been predicted to show Rashba-type SOC splitting [19], and Janus $WSiGeZ_4$ systems display symmetry-dependent Berry curvature and spin texture [20]. Magnetic $MA_2Z_4$ members provide complementary examples, including spin-valley-coupled $VSi_2N_4$ [21] and $CrSi_2N_4/CrSi_2P_4$, where SOC-driven valley spin splitting can be tuned by strain [22]. Taken together, these studies show that the valley response is controlled by the symmetry and orbital composition of the band edges, not by Janus asymmetry alone.

Cr-based septuple layers provide a particularly relevant reference point. First-principles calculations on $CrC_2N_4$ established a dynamically and thermally stable direct-gap semiconductor with high stiffness, carrier mobility, visible-light absorption, and piezoelectric response [23]. A later SOC-resolved study showed that the K/K′ band edges possess opposite out-of-plane spin character, with frontier states dominated by Cr-*d* and N-*p* hybridization, and reported strain-dependent changes in Berry curvature and optical response [24]. These results identify $CrC_2N_4$ as a useful parent system for examining how chemical differentiation of one outer sublayer modifies Cr-centered SOC and valley physics.

Janus $CrSiCN_4$ and $CrGeCN_4$ were subsequently proposed by replacing the C atoms on one side of $CrC_2N_4$ with Si or Ge. Their structural stability, direct-gap electronic structure, Cr/N-dominated frontier states, band-edge charge distributions, electrostatic-potential differences, work functions, carrier effective masses and mobilities, strain-dependent band edges, optical absorption, and photocatalytic band alignment have already been characterized [25]. The same study reported spin-degenerate band structures and PDOS and therefore omitted spin coupling from the subsequent calculations [25]. Consequently, the SOC-resolved band-edge electronic structure, the magnitude and orbital origin of spin splitting at the K/K′ valleys, the valley-resolved spin character, and the Berry-curvature response of these Janus monolayers were not addressed in that study. The evolution of these quantities and their response to Si-to-Ge substitution therefore remain to be clarified.

To address these research gaps, first-principles calculations were performed on pristine and biaxially strained Janus $CrSiCN_4$ and $CrGeCN_4$ monolayers. Their structural, electronic, piezoelectric, valleytronic, and optical properties were systematically investigated, including orbital contributions, Bader charge transfer, work function, Berry curvature, SOC-induced spin splitting, and carrier effective masses. Particular emphasis was placed on the interplay between frontier-state orbital character, valley-dependent Berry curvature, and strain-tunable electronic transitions.

## 2 Computational Methodology

First-principles calculations were performed using density functional theory (DFT) as implemented in the Vienna Ab-initio Simulation Package (VASP) [26,27]. The electron-ion interactions were described through the projector augmented-wave (PAW) method [27]. Exchange-correlation effects were treated using the Perdew-Burke-Ernzerhof (PBE) functional [28] within the generalized gradient approximation (GGA). The Heyd-Scuseria-Ernzerhof hybrid functional (HSE06) [29,30] was additionally employed to obtain more accurate electronic band gaps.

The monolayer was modeled under periodic boundary conditions with a vacuum region of approximately 25 Å along the surface normal. This separation minimized artificial coupling between periodic images. The Numerical convergence criteria was established by using a tested plane-wave kinetic-energy cutoff of 500 eV and the Brillouin zone sampling resolution of $2\pi \times$ 0.03 Å$^{-1}$. During relaxation, the atomic positions and in-plane lattice parameters were optimized, while the out-of-plane lattice parameter was fixed. A self-consistent dipole correction was applied at the center of the vacuum region to account for the intrinsic polarization and its associated electrostatic asymmetry. The van der Waals (vdW) interactions between layers were incorporated through the DFT-D3 dispersion correction to the total energy [31]. The convergence tolerance is $10^{-5}$ eV for total energy and 0.02 eV/Å for all forces. Charge redistribution was quantified using Bader charge analysis [32].

The response to biaxial strain was investigated by uniformly varying the lattice parameters in the xy-plane. The applied strain was defined as,

$$\varepsilon = \frac{a - a_0}{a_0} \times 100\%, \quad (1)$$

where $a_0$ and a are the equilibrium and strained lattice constants, respectively. For each applied strain the internal atomic coordinates were reoptimized while the lattice parameters were held at their prescribed values. Fully relativistic noncollinear calculations incorporating spin-orbit coupling were subsequently performed on the relaxed structures. Brillouin zone sampling employed a (24×24×1) k-point mesh for both the self-consistent calculations and the non-self-consistent calculations used for Wannier interpolation. Maximally localized Wannier functions were generated with Wannier90 [33] and used to evaluate Berry curvature and associated quantities in WannierBerri [34]. The elastic stiffness tensor and piezoelectric stress tensor are directly calculated by the DFPT method in the VASP package. The optical properties were derived from the frequency dependent complex dielectric function to examine the evolution of the optical response under strain. Since electron-hole interactions were not explicitly treated through the Bethe-Salpeter equation, the calculated spectra were used primarily to assess the evolution of the optical response under strain.

# 3 Results and Discussion

## 3.1 Structural Properties of Monolayer $CrMCN_4$ (M = Si/Ge)

The optimized geometries of monolayer $CrCSiN_4$ and $CrCGeN_4$ exhibit hexagonal in-plane lattices with seven atoms per primitive cell, arranged into seven distinct atomic layers along the out-of-plane direction (shown in **Figure 1**). Their stacking sequences are N-Si-N-Cr-N-C-N and N-Ge-N-Cr-N-C-N, respectively. In both structures, the central Cr layer is sandwiched between chemically inequivalent N-C-N and N-Si-N or N-Ge-N sublayers, giving rise to an asymmetric septuple-layer architecture. The optimized in-plane lattice constants are a = b = 2.689 Å for $CrCSiN_4$ and 2.743 Å for $CrCGeN_4$, with an angle of approximately 120° between the in-plane lattice vectors. Replacing Si with Ge therefore increases the equilibrium in-plane lattice constant by approximately 2.03%, while preserving the seven-layer stacking arrangement. This value is in excellent agreement with the previously reported value [25]. The increment in the lattice parameter results mainly from the larger ionic radius of Ge compared to Si. Previous investigations have confirmed the mechanical, dynamical, and thermodynamic stability of this structure. The calculated equilibrium structural parameters are summarized in **Table 1**.

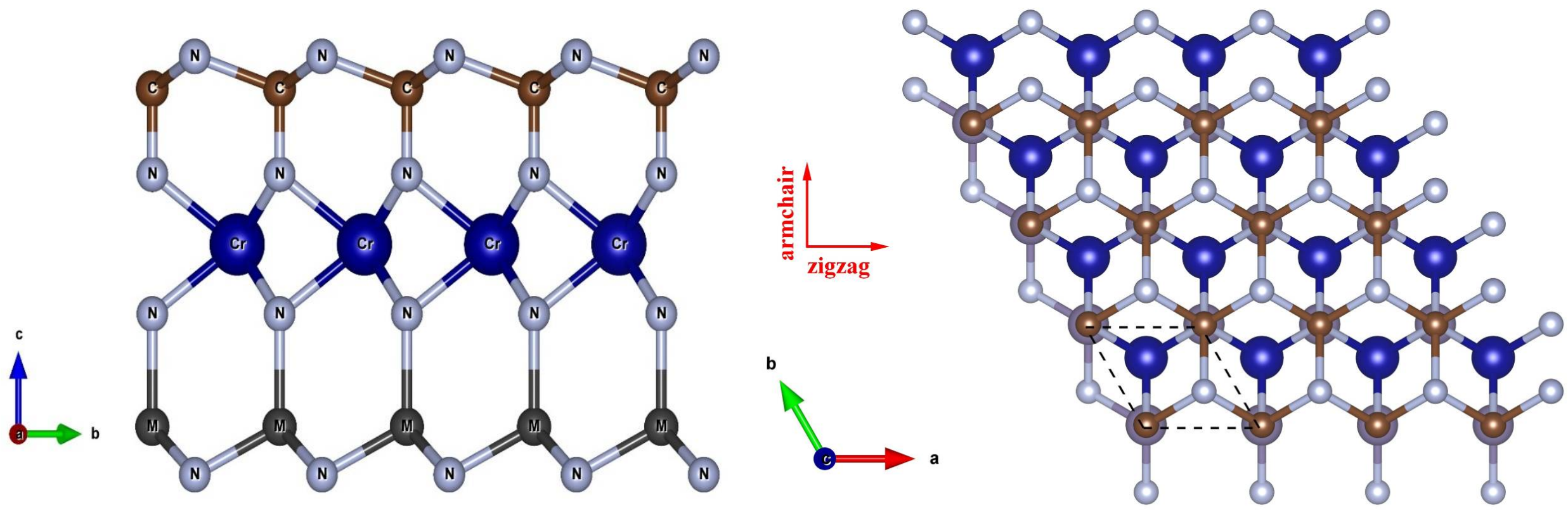

**Figure 1.** Optimized geometrical structures of monolayer $CrMCN_4$ (M = Si and Ge). The side view (left) and top view (right) depict the atomic configuration of the monolayers, showing the arrangement of Cr, C, N, and Si/Ge atoms. The corresponding unit cell and crystallographic directions are also indicated.

**Table 1.** Structural parameters of monolayer $CrSiCN_4$, $CrGeCN_4$ and $CrC_2N_4$. Here, $a$ is the in-plane lattice constant, $d_{Cr\text{-}N(in)}$ and $d_{Cr\text{-}N(out)}$ represent the two inequivalent Cr-N bond lengths, $d_{C\text{-}N(out)}$ and $d_{Si/Ge\text{-}N(out)}$ denote the corresponding C-N and Si/Ge-N bond lengths, respectively, and $t_{eff}$ is the effective thickness of the monolayers.

| Compound | $a$ (Å) | $d_{Cr\text{-}N(in)}$ (Å) | $d_{Cr\text{-}N(out)}$ (Å) | $d_{C\text{-}N(out)}$ (Å) | $d_{Si/Ge\text{-}N(out)}$ (Å) | $t_{eff}$ (Å) | Reference |
|---|---|---|---|---|---|---|---|
| $CrSiCN_4$ | 2.69 | 1.94 | 1.94 | 1.42 | 1.74 | 6.65 | This work |
| $CrGeCN_4$ | 2.74 | 1.96 | 1.96 | 1.41 | 1.86 | 6.92 | This work |
| $CrC_2N_4$ | 2.51 | 1.89 | 1.44 | 1.54 | - | 9.40 | [24] |
| $CrSiCN_4$ | 2.70 | - | - | - | - | - | [25] |
| $CrGeCN_4$ | 2.76 | - | - | - | - | - | [25] |

## 3.2 Charge Density and Bader Charge Analysis

The band-decomposed charge densities of the valence band maximum (VBM) and conduction band minimum (CBM) for $CrSiCN_4$ and $CrGeCN_4$ are presented in **Figure 2**. The band-edge charge distributions are primarily localized around the Cr atoms and their neighboring N atoms, whereas the contribution from C atoms is comparatively negligible. This observation is consistent with the orbital-resolved electronic structures, confirming that the major bonds are mainly governed by Cr-d and N-p orbital interactions, as discussed in detail in the following section.

To further elucidate the charge redistribution and bonding characteristics, Bader charge analysis was performed using the converged charge density. The results reveal substantial charge transfer from Si/Ge, Cr, and C atoms toward the more electronegative N atoms. For $CrSiCN_4$, the Si, Cr, and C atoms lose 2.217e, 1.374e, and 0.697e, respectively, whereas the outer and inner N atoms accumulate 2.012e and 2.378e, respectively. The larger charge accumulation on the inner N atoms indicates their stronger involvement in the Cr-N bonding network. A similar charge-transfer behavior is observed in $CrGeCN_4$. These results demonstrate that both monolayers possess pronounced charge polarization with mixed ionic-covalent bonding characteristics, primarily originating from electron redistribution between electropositive Si/Ge, Cr, C atoms and electronegative N atoms.

## 3.3 Electronic Properties of the $CrMCN_4$ (M = Si/Ge) Monolayer

The electronic structures of monolayer $CrMCN_4$ (M = Si/Ge) were investigated using HSE06 band structures (without SOC), projected density of states (PDOS), PBE+SOC band structure with spin expectation value along the z direction and orbital-resolved band dispersions, as presented in **Figure 2-4**.

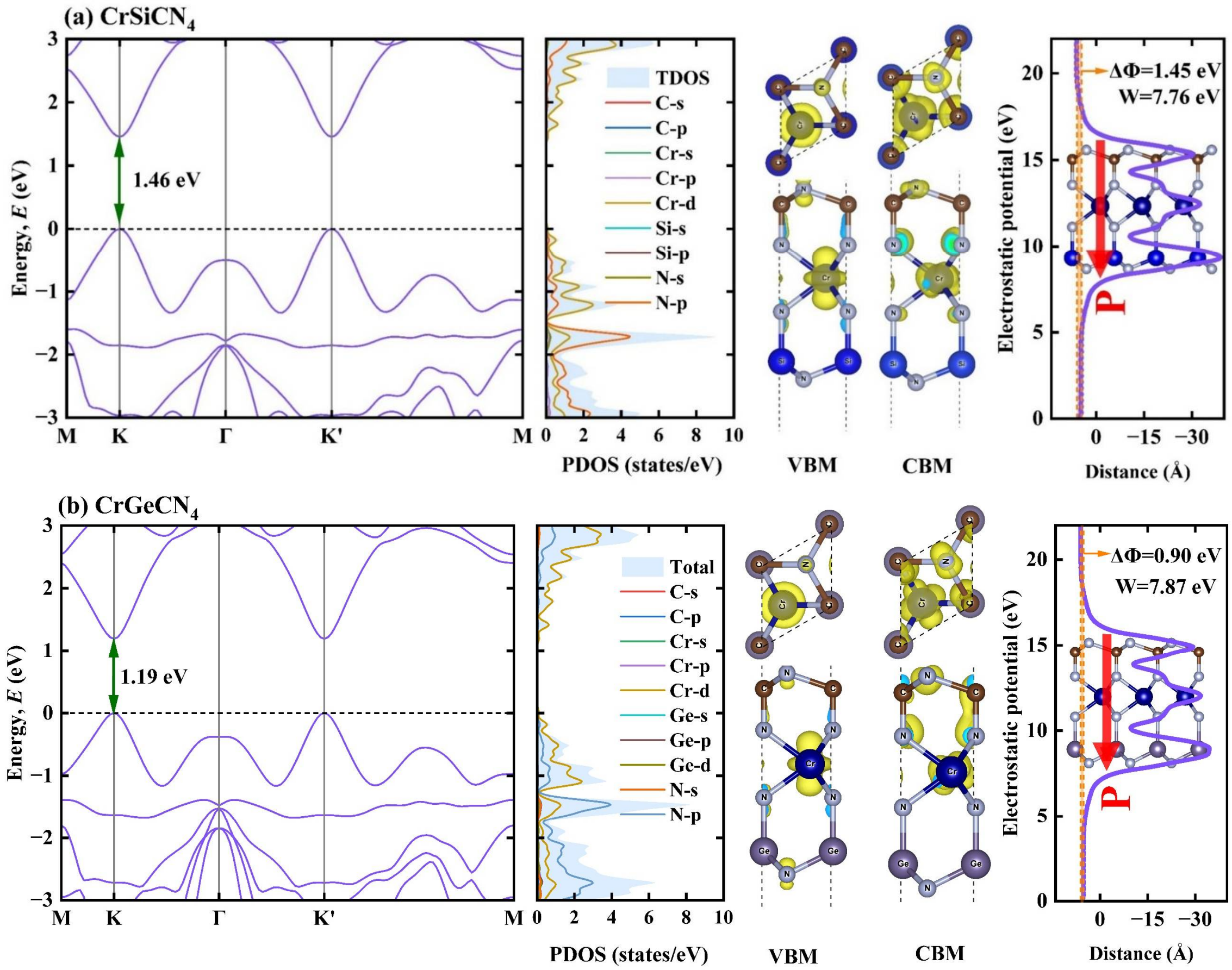

**Figure 2.** Electronic properties of monolayer $CrSiCN_4$ and $CrGeCN_4$ calculated using the HSE06 functional without spin-orbit coupling (SOC). (a, b) Electronic band structures of $CrSiCN_4$ and $CrGeCN_4$, respectively. The corresponding projected density of states (PDOS) is shown alongside the band structures. The band-decomposed charge densities of the valence band maximum (VBM) and conduction band minimum (CBM), together with the planar-averaged electrostatic potential distribution along the out-of-plane (Z) direction, are also presented.

The electronic structures of monolayer $CrSiCN_4$ and $CrGeCN_4$ were evaluated using the HSE06 functional without spin-orbit coupling (SOC), as illustrated in **Figure 2**. Both monolayers exhibit direct band gap semiconducting behavior, with calculated band gaps of 1.46 eV and 1.19 eV for $CrSiCN_4$ and $CrGeCN_4$, respectively. The valence band maximum (VBM) and conduction band minimum (CBM) are located at the symmetry equivalent K and K′ valleys, revealing the intrinsic valley characteristics of these systems. The surface electronic properties were further elucidated through the planar-averaged electrostatic potential profiles ( **Figure 2(a, b)**). The work functions, obtained from the energy separation between the vacuum level and the Fermi level, are calculated to be 7.76 eV and 7.87 eV for $CrSiCN_4$ and $CrGeCN_4$, respectively. These large work functions indicate strong surface electron binding and suggest favorable interfacial compatibility with appropriate electrode materials, which is crucial for controlling carrier injection and band alignment in emerging valleytronic, optoelectronic, and photocatalytic applications. Notably, the electrostatic potential profiles display a clear asymmetry between the two surfaces, originating from the intrinsic out-of-plane polarization. This polarization-induced dipole generates an internal electric field across the monolayers and leads to unequal vacuum potentials on opposite surfaces. The resulting surface potential differences are calculated as $\Delta\Phi = 1.45$ eV for $CrSiCN_4$ and 0.90 eV for $CrGeCN_4$, respectively, confirming the presence of a strong polarization-driven electrostatic imbalance.

The spin-orbit-coupled (SOC) induced band structures were further analyzed to explore the valleytronic properties, as shown in **Figure 3.** The SOC-inclusive band structures reveal reduced band gaps of 1.23 eV and 1.09 eV for $CrSiCN_4$ and $CrGeCN_4$, respectively, calculated using the PBE functional. These values are slightly smaller than the corresponding HSE06 results, consistent with the well-known tendency of the PBE functional to underestimate semiconductor band gaps due to the absence of exact exchange contributions. Despite this quantitative deviation, the PBE+SOC calculations successfully reproduce the essential electronic features and preserve the fundamental band characteristics of both monolayers. Therefore, considering the substantial computational cost of SOC calculations, the PBE functional was employed for subsequent SOC-dependent analyses, providing an effective balance between computational efficiency and reliable description of the relevant electronic properties.

The spin-polarized band structures and projected density of states (PDOS) without SOC are provided in the Appendix. The complete overlap between spin-up and spin-down channels confirms the nonmagnetic ground state of both monolayers. Upon inclusion of SOC, the semiconducting nature is preserved, while the spin degeneracy of the valley-edge states is lifted. As shown in **Figure 3**, the spin expectation values reveal that the band-edge states at the K and K′ valleys are dominated by the out-of-plane spin component ($S_z$), with negligible in-plane contributions. The K valley states exhibit predominantly spin-down polarization ($\langle S_z \rangle \approx -0.5$), whereas the time-reversal related K′ valley states possess opposite spin polarization ($\langle S_z \rangle \approx +0.5$). The nearly degenerate K/K′ valley energies are consistent with preserved time-reversal symmetry in these nonmagnetic systems and this small deviation can be associated with the moderate SOC strength of Cr-derived states compared with heavier transition-metal systems. Nevertheless, strong SOC-induced spin splitting occurs at the band edges, with valence-band splittings ($\Delta_v$) of 53.3 meV and 52.4 meV for $CrSiCN_4$ and $CrGeCN_4$, respectively, while the conduction-band splitting ($\Delta_c$) remains approximately 1.6 meV. The opposite spin polarization between the K and K′ valleys establish robust spin-valley locking, providing the basis for valley-selective optical transitions and Berry-curvature driven valley Hall phenomena. Both optimized $CrSiCN_4$ and $CrGeCN_4$ monolayers are non-centrosymmetric, although their upper and lower sublayers exhibit an approximate horizontal mirror relationship. The absence of spatial inversion symmetry originates from the inequivalent atomic configurations, where inversion operation does not map each atomic site onto an equivalent site occupied by the same element. Together with SOC and preserved time-reversal symmetry, this broken inversion symmetry enables spin splitting of the valley-edge states.

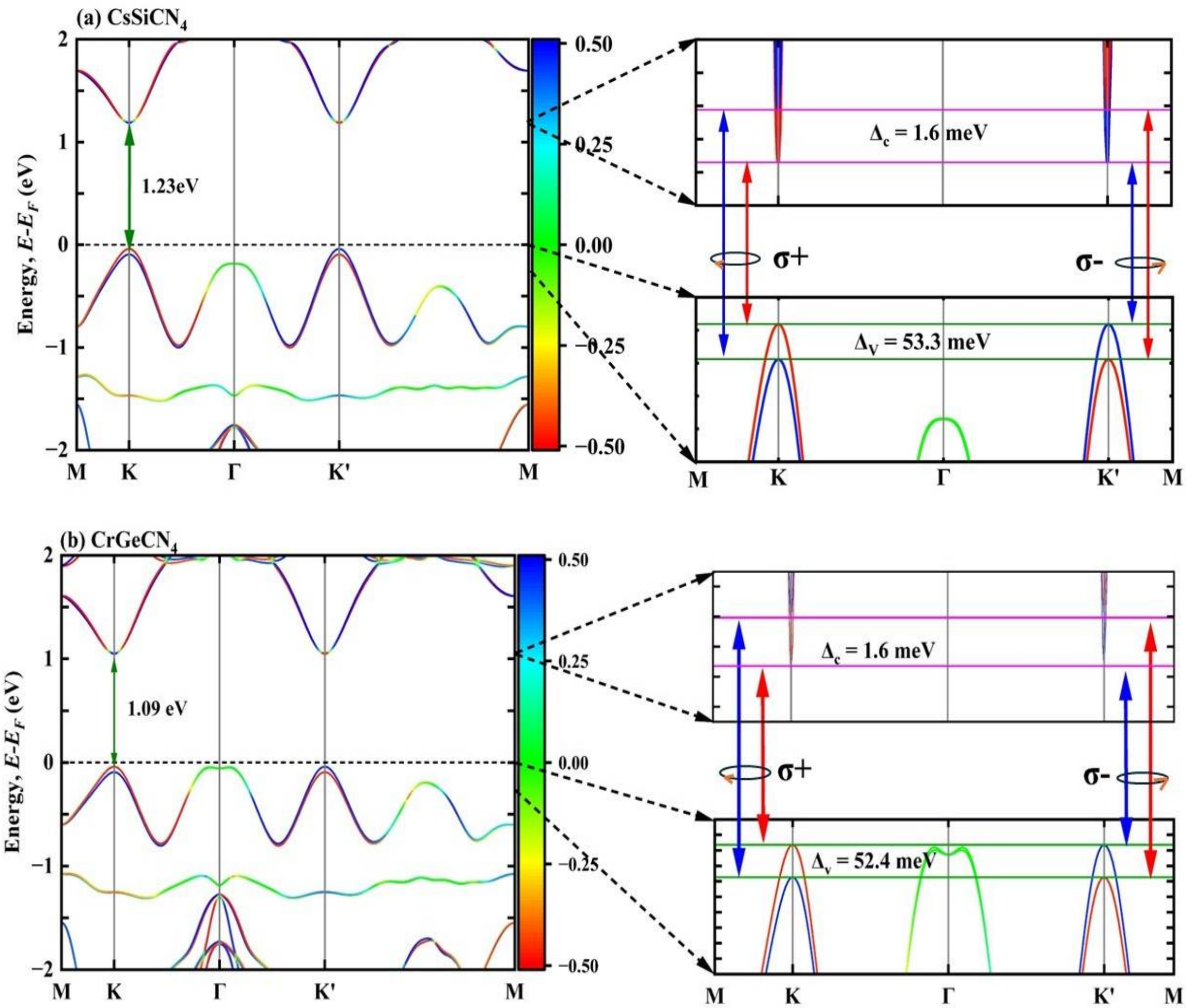

**Figure 3.** SOC-induced spin splitting and valley-dependent electronic structures of monolayer (a) $CrSiCN_4$ and (b) $CrGeCN_4$ obtained using PBE+SOC calculations. The band gaps are marked by blue arrows, while the enlarged K/K′ valley regions show the conduction band ($\Delta_c$) and valence band ($\Delta_v$) spin splittings. Red and blue arrows indicate valley-selective optical transitions coupled with right and left-handed circularly polarized light, respectively. The color map represents the spin expectation value along the z direction ($\langle S_z \rangle$) of the spinor states.

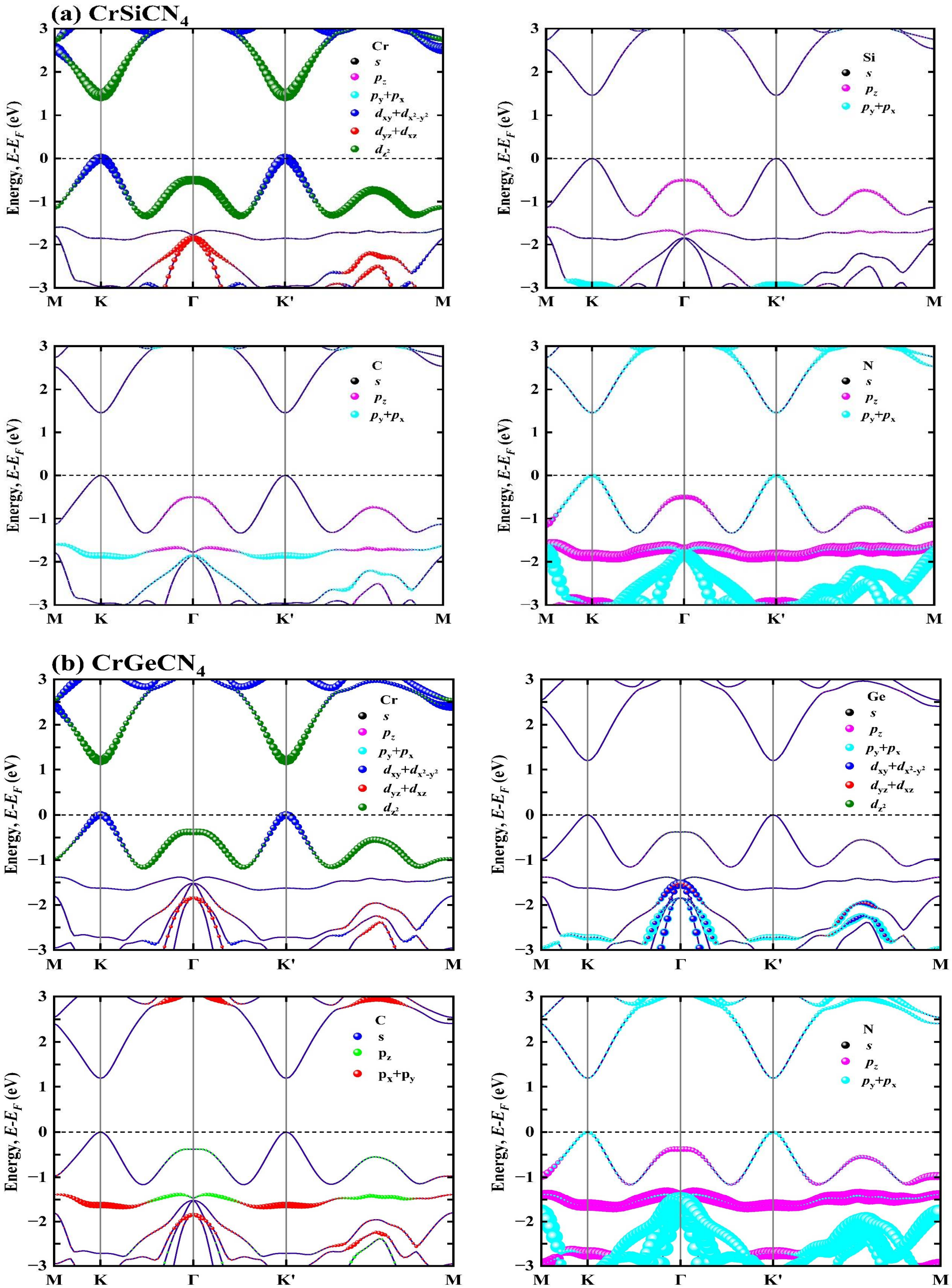

**Figure 4.** Orbital-projected band structures of monolayer (a)$CrSiCN_4$ and (b)$CrGeCN_4$, illustrating the orbital contributions of Cr-*d*, Si/Ge-*p*, C-*p*, and N-*p* states near the band edges.

The orbital-resolved projected density of states (PDOS) was further analyzed to elucidate the orbital contributions governing the electronic structures of $CrSiCN_4$ and $CrGeCN_4$, as presented in **Figure 2**. The PDOS profiles reveal that the band-edge states of both monolayers are primarily derived from Cr-*d* orbitals with significant hybridization from N-*p* states, whereas the Si and C derived states contribute minimally near the band edge. This indicates that the electronic properties around the Fermi level and the low-energy excitations of both systems are mainly governed by Cr-N orbital hybridization.

Further insight into the orbital character of the band-edge states is provided by the orbital-projected band structures in **Figure 4(a, b)**. The Si/Ge- and C-derived states contribute only weakly near the band edges, whereas pronounced Cr-d and N-p contributions are observed around the K/K′ valleys. Specifically, the VBM is predominantly derived from the in-plane Cr-$d_{xy}$ and $d_{x^2-y^2}$orbitals, strongly hybridized with N-$p_x/p_y$ and N- $p_z$ states, while the CBM is mainly characterized by Cr-$d_{z^2}$ states with appreciable N-$p_x/p_y$ hybridization. This distinct orbital composition establishes Cr-N hybridization as the microscopic origin of the valley-edge states and underpins the SOC-driven spin-valley characteristics of both monolayers.

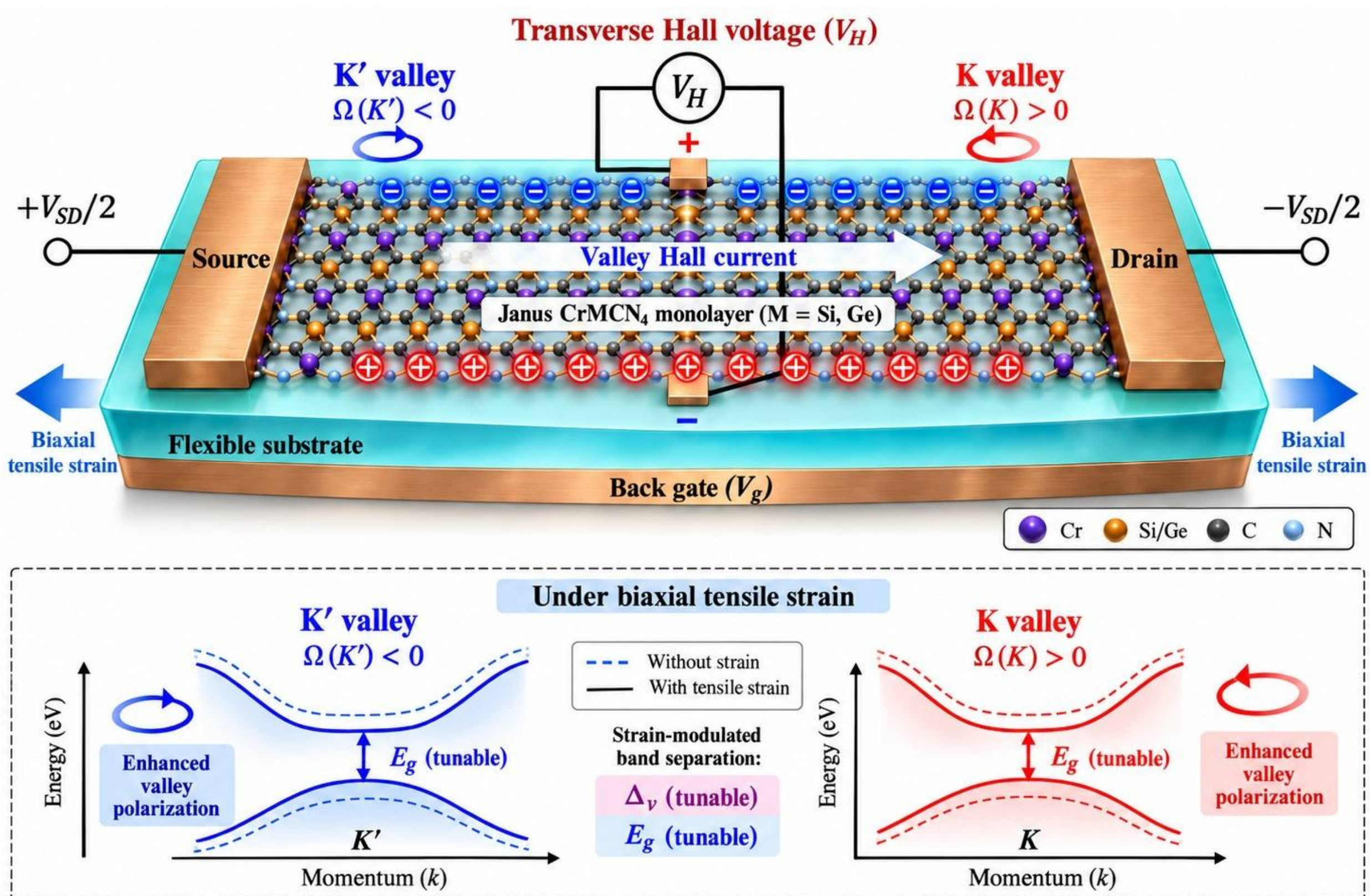

**Figure 5. Schematic illustration of a strain-tunable spin-valley / valley Hall device based on Janus $CrMCN_4$ (M =Si, Ge).** The monolayer channel is integrated on a flexible substrate between source and drain electrodes with back gate. Biaxial tensile strain modulates the electronic structure and enhances transverse Hall-voltage response. Charge accumulation at opposite channel edges produce a measurable Hall voltage $V_H$.

### 3.3.1 Comparative Analysis

To assess the valleytronic characteristics of monolayer $CrSiCN_4$ and $CrGeCN_4$, the SOC-induced spin splittings at the band edges are benchmarked against representative two-dimensional valleytronic materials, as summarized in **Table 2**. Here, $\Delta_v$ and $\Delta_c$ denotes the spin splitting magnitudes at the valence and conduction band edges, respectively. Both monolayers exhibit pronounced band-edge spin-splitting asymmetry, with valence-band splittings of 53.3 meV and 52.4 meV for $CrSiCN_4$ and $CrGeCN_4$, respectively, accompanied by much smaller conduction-band splittings of 1.6 meV. The obtained valence-band splittings are comparable to several reported valleytronic systems, including $CrC_2N_4$, $VSi_2P_4$, $VBNS_2$, and $WS_2/MnO_2$, while remaining lower than the larger splittings reported for $CrSi_2N_4$, $MoSi_2N_4$, and $CrSi_2P_4$, which originate from their stronger SOC interactions.

**Table 2.** Comparison of valley related band-edge splitting in $CrSiCN_4$, $CrGeCN_4$, and selected representative 2D valleytronic systems. The values $\Delta_v$ and $\Delta_c$ are reported in meV. For nonmagnetic systems, these quantities represent SOC-induced valley-edge spin splitting, whereas for magnetic or proximity-coupled systems, they correspond to K/K′ valley splitting or valley polarization.

| Compound | $\Delta_v$ (meV) | $\Delta_c$ (meV) | Reference |
|---|---|---|---|
| $CrSiCN_4$ | 53.3 | 1.6 | This work |
| $CrGeCN_4$ | 52.4 | 1.6 | This work |
| $CrC_2N_4$ | 51.9 | 1.7 | [24] |
| $CrSi_2N_4$ | 130 | 26 | [22] |
| $MoSi_2N_4$ | 130-172 | 3-17 | [16] |
| $CrSi_2P_4$ | 170 | 23 | [22] |
| $VSi_2P_4$ | 49.4 | 3 | [35] |
| $VBNS_2$ | 48.6 | - | [36] |
| $WS_2/MnO_2$ | 43 | - | [37] |

The pronounced $\Delta_v/\Delta_c$ asymmetry arises from the orbital-dependent SOC response of the band-edge states. The VBM dominated by Cr-$d_{xy}/d_{x^2-y^2}$ orbitals strongly hybridized with N-$p_{xy}$ states which exhibits enhanced SOC-induced splitting, whereas the Cr-$d_{z^2}$ dominated CBM shows comparatively weaker splitting. Therefore, $CrSiCN_4$ and $CrGeCN_4$ represent promising nonmagnetic valleytronic platforms by combining substantial band-edge spin splitting, robust spin-valley locking, and valley-contrasting Berry curvature, highlighting their suitability for valley-dependent electronic and optoelectronic applications. Building on these intrinsic properties, **Figure 5** illustrates a prototype strain-controlled valley Hall device concept for tuning valley-dependent transport.

### 3.4 Valley Physics: Berry Curvature

The Berry curvature serves as a momentum-space geometric descriptor of electronic Bloch states and acts as an effective magnetic field in reciprocal space, governing anomalous carrier dynamics and valley-dependent transport phenomena. In time-reversal-symmetric systems, the Berry curvature at the K and K′ valleys exhibit opposite signs, resulting in valley-contrasting electronic responses. Following the valley effective model proposed by Li *et al.* [**16**], the low-energy electronic states near the inequivalent K/K′ valleys can be described within a two-band Dirac Hamiltonian framework. Without considering spin-orbit coupling (SOC), the Hamiltonian around the valley index $\tau = \pm 1$ is expressed as:

$$H_0^\tau = \alpha\left(\tau k_x \sigma_x + k_y \sigma_y\right) + \frac{\Delta}{2}\sigma_z, \tag{2}$$

Here, $k_i$ denotes the wave vector measured relative to the valley center, $\sigma_i$ represents the Pauli matrices in the two-orbital basis, $\alpha$ is the effective coupling coefficient and $\Delta$ corresponds to the band-gap parameter. Upon incorporating spin-orbit coupling (SOC), the effective Hamiltonian is modified as:

$$H^\tau = \alpha\left(\tau k_x \sigma_x + k_y \sigma_y\right) + \frac{\Delta}{2}\sigma_z - \frac{\lambda}{2}\tau(\sigma_z - 1)\hat{s}_z, \tag{3}$$

Here, $\lambda$ represents the effective SOC strength, and $\hat{s}_z$ denotes the spin Pauli matrices. This Hamiltonian captures the essential spin-valley-coupled nature of the band-edge states, where SOC induces pronounce splitting in the valence band while producing comparatively weaker splitting in the conduction band. For Berry curvature, the following scheme can be used:

$$\Omega_{n,z}(\mathrm{k}) = -2\,\mathrm{Im}\sum_{m\neq n}\frac{\langle n\mathbf{k}|\hat{v}_x|m\mathbf{k}\rangle\langle m\mathbf{k}|\hat{v}_y|n\mathbf{k}\rangle}{(\omega_m - \omega_n)^2}, \tag{4}$$

In this expression, $\hat{v}_x$ and $\hat{v}_y$ correspond to the velocity operators along the in-plane directions, while $E_n = \hbar\omega_n$ denotes the eigenvalue associated with the Bloch state $|n\mathbf{k}\rangle$. These velocity operators are evaluated directly from the effective Hamiltonian through:

$$\hat{v}_i = \frac{1}{\hbar}\frac{\partial H^\tau}{\partial k_i}, \quad i = x, y \tag{5}$$

Incorporating the SOC contribution into the effective Hamiltonian (**Equation 2**), the Berry curvature associated with the conduction band for a given valley index $\tau$ and spin state $s = \pm 1$ can be expressed as [**16**]:

$$\Omega_c^\tau(\mathbf{k}, s) = -\tau\frac{2\alpha^2\tilde{\Delta}_{\tau s}}{\left(\tilde{\Delta}_{\tau s}^2 + 4\alpha^2 k^2\right)^{3/2}}, \tag{6}$$

here

$$\tilde{\Delta}_{\tau s} = \Delta - \tau s\lambda, \tag{7}$$

For the corresponding spin-split valence band, the Berry curvature exhibits an opposite sign,

$$\Omega_v^\tau(\mathbf{k}, s) = -\Omega_c^\tau(\mathbf{k}, s). \tag{8}$$

**Equation 5** demonstrates that the Berry curvature reverses its sign between the two time-reversal related K and K′ valleys due to the valley dependent factor τ. Consequently, the K and K′ valleys exhibit finite Berry curvatures of opposite signs, with the maximum magnitude occurring at the valley center (k =0).

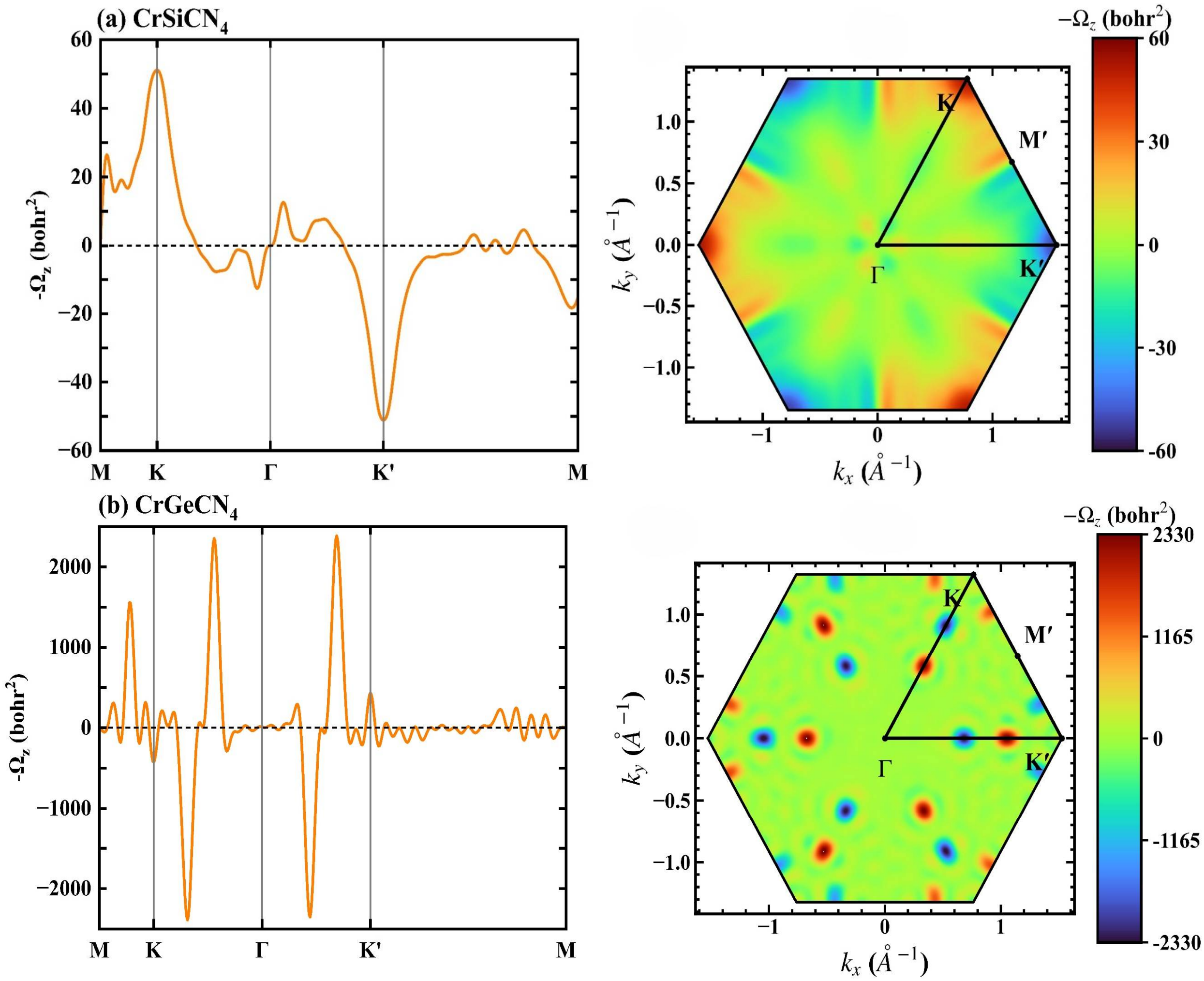

**Figure 6.** Berry curvature, $-\Omega_z(\text{bohr}^2)$, of monolayer (a) $CrSiCN_4$ and (b) $CrGeCN_4$ along the high-symmetry path $M \rightarrow K \rightarrow \Gamma \rightarrow K' \rightarrow M$, showing opposite Berry curvature at the K and K′ valleys. The corresponding contour maps reveal the valley-contrasting Berry curvature distribution throughout the hexagonal Brillouin zone.

Under an applied in-plane electric field, the finite Berry curvature induces an anomalous transverse velocity, giving rise to Hall-type transport phenomena such as the valley Hall effect. Since $CrSiCN_4$ and $CrGeCN_4$ preserve time-reversal symmetry, the net charge Hall conductivity vanishes when the K and K′ valleys are equally occupied. However, the opposite Berry curvatures at these time-reversal related valleys drive carriers toward opposite transverse directions, resulting in a pure valley Hall current. This valley-contrasting response can be expressed as:

$$\Omega_z(K) = -\Omega_z(K') \tag{9}$$

Therefore, despite the absence of a net charge Hall current, the opposite anomalous velocities generated in the K and K′ valleys produce a finite valley-polarized Hall response.

The Berry curvature of monolayer $CrSiCN_4$ and $CrGeCN_4$ exhibits strong valley-contrasting behavior, satisfying $\Omega_z(K) = -\Omega_z(K')$ due to time-reversal symmetry. This opposite Berry curvature at the time-reversal-related valleys produces anomalous transverse carrier motion under an in-plane electric field, described by $v_a = -\frac{e}{\hbar}\mathbf{E} \times \mathbf{\Omega}(\mathbf{k})$, giving rise to valley-dependent Hall transport. Although the charge Hall conductivity cancels when the K and K′ valleys are equally populated, the opposite anomalous velocities of carriers from the two valleys generate a finite valley Hall response. Such valley-contrasting Berry curvature and valley Hall physics have been widely established in inversion-asymmetric two-dimensional systems [**4**,**38**,**39**].

For $CrSiCN_4$, the Berry curvature is mainly concentrated around the K/K′ valleys with opposite signs, reflecting the intrinsic valley-dependent electronic structure. In contrast, $CrGeCN_4$ exhibits a strongly enhanced Berry-curvature landscape with sharp hot spots reaching the order of $10^3$ $bohr^2$, significantly exceeding the response of $CrSiCN_4$. This enhancement cannot be attributed solely to stronger valley-edge SOC because both systems exhibit nearly identical SOC-induced spin splittings. Furthermore, the reduction of the fundamental band gap provides only a limited enhancement within the massive Dirac picture. Instead, the giant Berry curvature in $CrGeCN_4$ is attributed to multiband electronic reconstruction induced by Ge substitution, including modified orbital hybridization, band alignment, and reduced interband energy separation.

According to the Kubo formulation, Berry curvature depends on interband velocity matrix elements and the inverse square of energy separation (**Equation 4**). Therefore, small energy separations or SOC-mediated avoided crossings can strongly amplify the Berry-curvature response, as shown in **Figure 3(b)** near Γ point. Similar interband-driven enhancement of Berry curvature and Hall-related phenomena has been reported in multiband electronic systems [**40**–**42**]. Hence, the unusually large Berry-curvature hot spots in $CrGeCN_4$ originate primarily from enhanced multiband coupling rather than a simple increase of intrinsic valley curvature.

## 3.5 Photocatalytic Properties

Among the two monolayers, $CrSiCN_4$ exhibits a band gap exceeding the thermodynamic minimum of 1.23 eV required for overall water splitting. Its photocatalytic activity was assessed by aligning the band edges with the water redox potentials, where the CBM and VBM must provide sufficient driving forces for HER and OER, respectively.

Owing to its Janus asymmetry, $CrSiCN_4$ exhibits an intrinsic electrostatic potential difference of 1.48 eV, generating a built-in electric field from the lower to the upper surface (**Figure 7**). This field promotes spatial separation of photogenerated carriers, driving electrons toward the lower surface for HER and holes toward the upper surface for OER [**43**] . Accordingly, the $H^+/H_2$ and $O_2/H_2O$ redox potentials were referenced to the corresponding surface-specific vacuum levels. The redox potentials of water were referenced to the vacuum level using the following relations [**44**]:

$$V_{H^+/H_2} = E_{vacuum} - 4.44 + pH \times 0.059 \text{ eV} \qquad (10)$$

$$V_{O_2/H_2O} = E_{vacuum} - 5.67 + pH \times 0.059 \text{ eV} \tag{11}$$

The band-edge alignment satisfies the OER requirement but provides insufficient driving force for HER, indicating that further band-edge modulation is necessary. Strain engineering therefore offers a promising route to optimize the band alignment and enable overall water splitting.

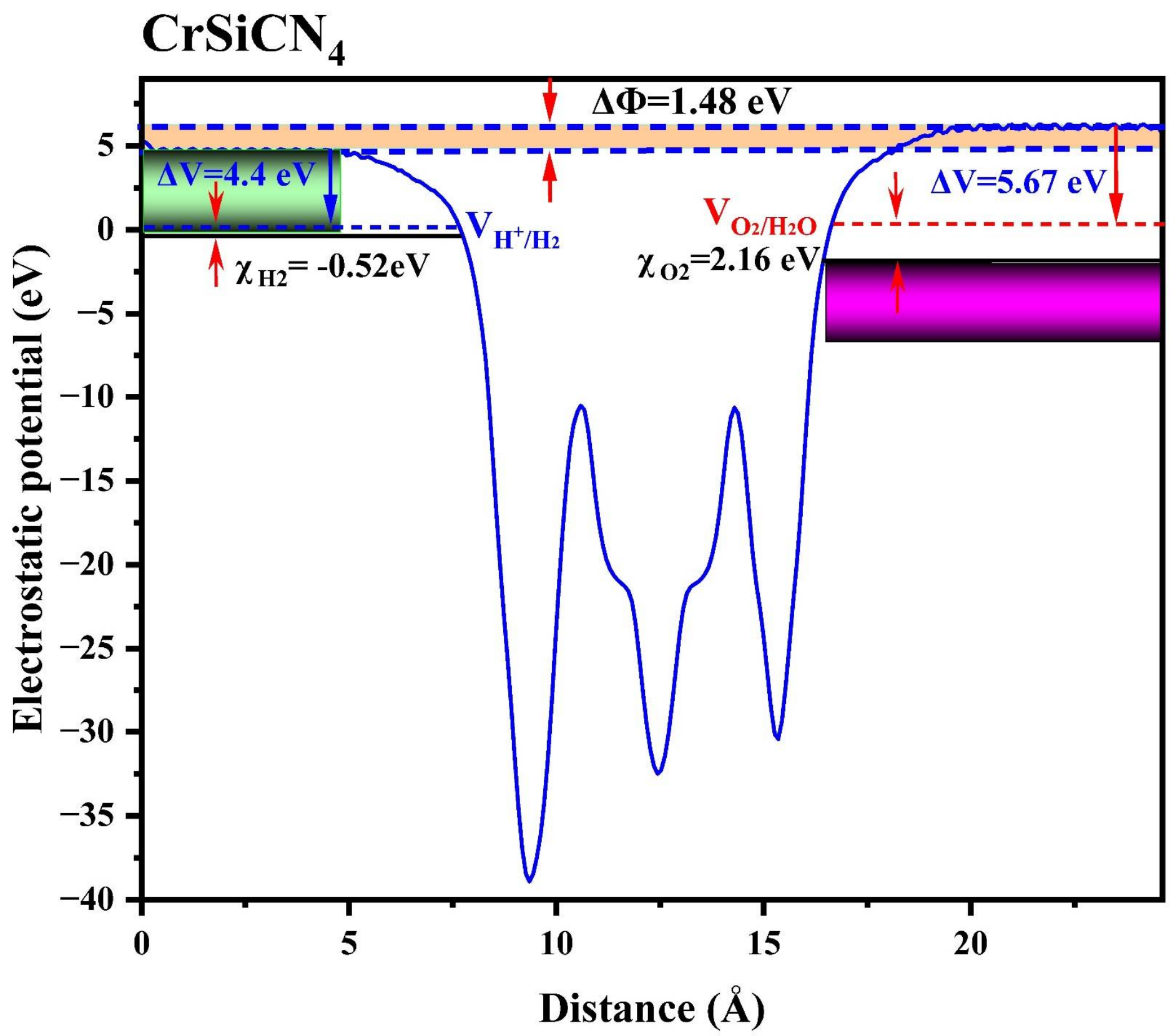

**Figure 7. Potential profiles.** The blue and red dashed lines represent the $H_2$ reduction and $H_2O$ oxidation potentials, while the green and magenta regions indicate the CBM and VBM positions, respectively. $\chi(H_2)$ and $\chi(O_2)$ denote the HER and OER driving forces. ΔΦ refers to the vacuum-level difference between the two surfaces.

## 3.6 Piezoelectric Properties

The piezoelectric behavior of both Janus monolayers is further examined by considering their in-plane and out-of-plane responses. Owing to the *P*3*m*1 symmetry for 2D, the elastic and piezoelectric tensors can be expressed in the following reduced forms [45,46]:

$$C = \begin{pmatrix} C_{11} & C_{12} & 0 \\ C_{12} & C_{11} & 0 \\ 0 & 0 & \frac{C_{11} - C_{12}}{2} \end{pmatrix}, \tag{12}$$

$$e = \begin{pmatrix} e_{11} & -e_{11} & 0 \\ 0 & 0 & -e_{11} \\ e_{31} & e_{31} & 0 \end{pmatrix}, \tag{13}$$

$$d = \begin{pmatrix} d_{11} & -d_{11} & 0 \\ 0 & 0 & -2d_{11} \\ d_{31} & d_{31} & 0 \end{pmatrix}. \tag{14}$$

The independent $d_{11}$ and $d_{31}$ coefficients can be derived from the relation $e_{ij} = d_{ij}C_{ij}$ as follows:

$$d_{11} = \frac{e_{11}}{C_{11} - C_{12}} \; and \; d_{31} = \frac{e_{31}}{C_{11} + C_{12}} \, . \tag{15}$$

As HSE and PBE yield similar electronic characteristics, apart from band-gap differences, the PBE functional was adopted for computational efficiency. Using the orthorhombic supercell (**Figure 1**), the piezoelectric coefficients were calculated from **Equation (12)-(15)** and are summarized in **Table 3**. The broken inversion symmetry of Janus $CrSiCN_4$ and $CrGeCN_4$ gives rise to both in-plane and out-of-plane piezoelectric responses. As shown in **Figure 8**, the calculated $e_{11}$ values are 11.157 and 13.136 × $10^{-10}$ C $m^{-1}$ for $CrSiCN_4$ and $CrGeCN_4$, respectively, with the electronic contribution dominating the in-plane response. These values are considerably larger than those reported for $MoBNTe_2$ (3.008 × $10^{-10}$ C $m^{-1}$), $MoSiGeN_4$ (5.116 × $10^{-10}$ C $m^{-1}$), and $MoSiN_3H$ (4.146 × $10^{-10}$ C $m^{-1}$), highlighting the pronounced in-plane piezoelectricity of the present monolayers. In contrast, the electronic and ionic contributions to $e_{31}$ have opposite signs and partially compensate each other, resulting in comparatively weaker out-of-plane responses.

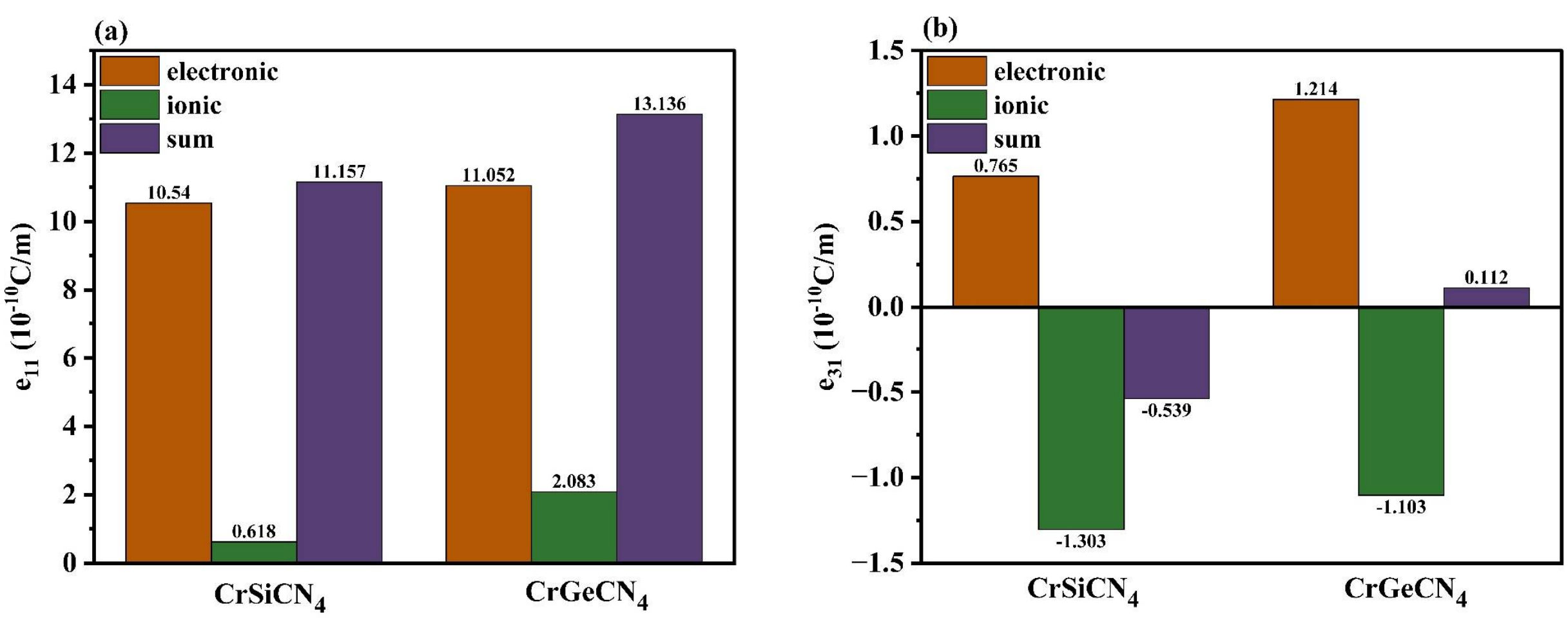

**Figure 8.** Piezoelectric stress coefficients (a) $e_{11}$ and (b) $e_{31}$ of Janus $CrSiCN_4$ and $CrGeCN_4$ monolayers, showing the electronic, ionic, and total contributions.

The corresponding piezoelectric strain coefficients $d_{11}$ are 3.305 and 5.084 pm $V^{-1}$ for $CrSiCN_4$ and $CrGeCN_4$, respectively. This demonstrates the stronger in-plane electromechanical response of $CrGeCN_4$. Conversely, $CrSiCN_4$ exhibits a larger $d_{31}$ of 0.082 pm $V^{-1}$ compared with 0.020 pm $V^{-1}$ for $CrGeCN_4$. Notably, the $d_{31}$ value of $CrSiCN_4$ exceeds those reported for $MoBNTe_2$ (0.062 pm $V^{-1}$), $MoSiGeN_4$ (0.014 pm $V^{-1}$), and $MoSiN_3H$ (0.058 pm $V^{-1}$), indicating a competitive vertical piezoelectric response. Overall, the large in-plane response of $CrGeCN_4$ and enhanced vertical response of $CrSiCN_4$ demonstrate complementary piezoelectric characteristics, making these Janus monolayers promising candidates for nanoscale electromechanical and energy-harvesting applications.

**Table 3.** Piezoelectric coefficients $e_{11}/e_{31}$ ($10^{-10}$ C $m^{-1}$) and $d_{11}/d_{31}$ (pm $V^{-1}$) of $CrMCN_4$ (M = Si/Ge).

| Material | $e_{11}$ | $d_{11}$ | $e_{31}$ | $d_{31}$ | Reference |
|---|---|---|---|---|---|
| $CrSiCN_4$ | 11.157 | 3.305 | 0.539 | 0.082 | This work |
| $CrGeCN_4$ | 13.136 | 5.084 | 0.112 | 0.020 | This work |
| $MoBNTe_2$ | 3.008 | 4.754 | 0.188 | 0.062 | [47] |
| $MoSiGeN_4$ | 5.116 | 1.494 | 0.087 | 0.014 | [19] |
| $MoSiN_3H$ | 4.146 | 1.622 | 0.290 | 0.058 | [48] |

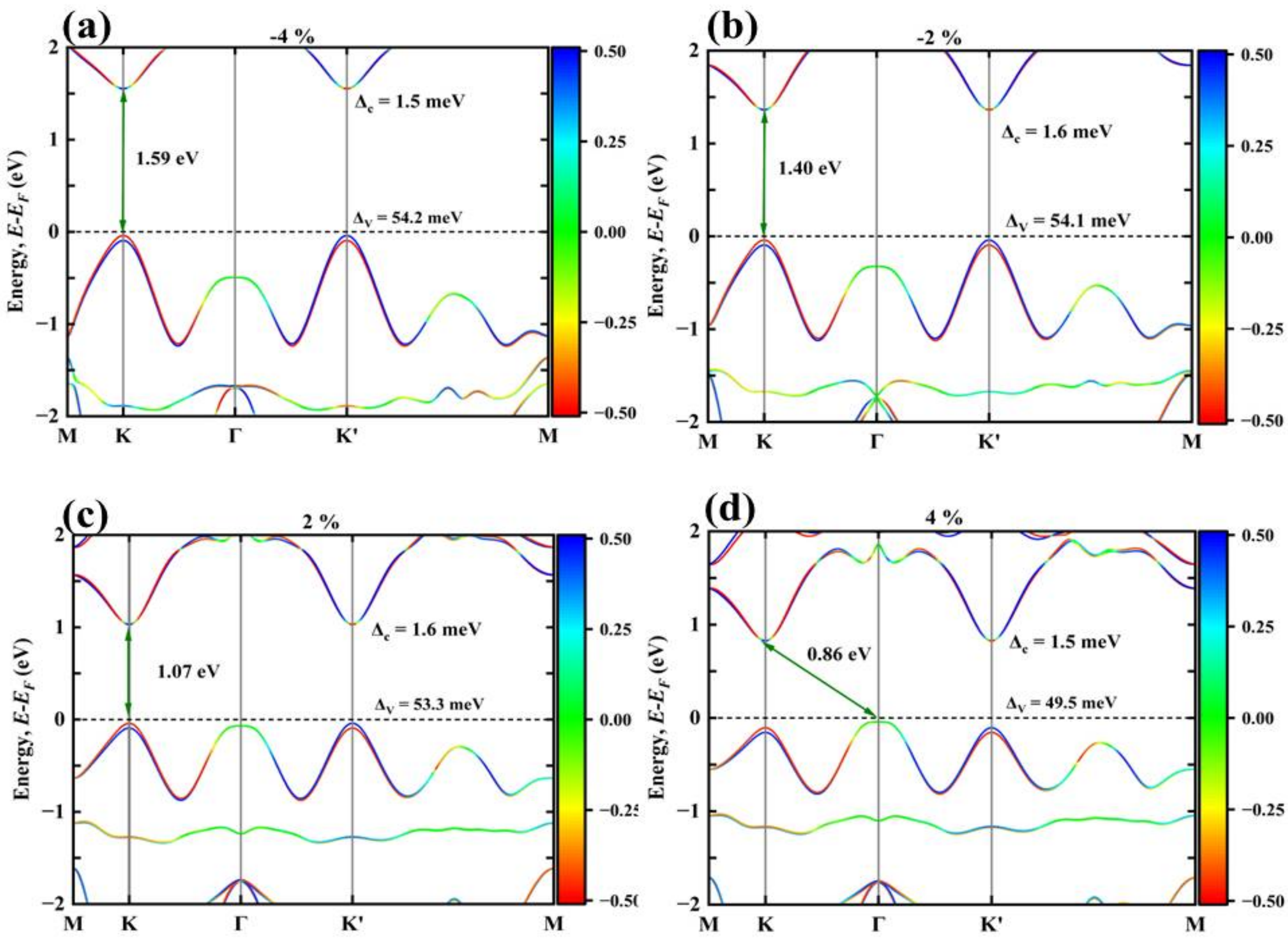

**Figure 9.** Spin-resolved band structures of monolayer $CrSiCN_4$ with SOC under biaxial strains of (a) −4%, (b) −2%, (c) +2%, and (d) +4%. The color scale represents the out-of-plane spin expectation value $\langle \mathbf{S_z} \rangle$, indicating strain-induced spin splitting and valley polarization.

### 3.7 Effect of Biaxial Strain on Band Structure, Spin Splitting

The strain-dependent electronic and spin properties of monolayer $CrSiCN_4$ and $CrGeCN_4$ were systematically explored using SOC inclusive spin-resolved band structure calculations. The results reveal that biaxial strain provides an effective means to simultaneously modulate the band topology, valley-dependent spin characteristics, and carrier transport properties. As illustrated in **Figure 9(a–d)** and **Figure 10(a–d)**, lattice deformation induces substantial shifts of the conduction-band minimum (CBM) and valence-band maximum (VBM), resulting in pronounced strain-driven electronic reconstruction.

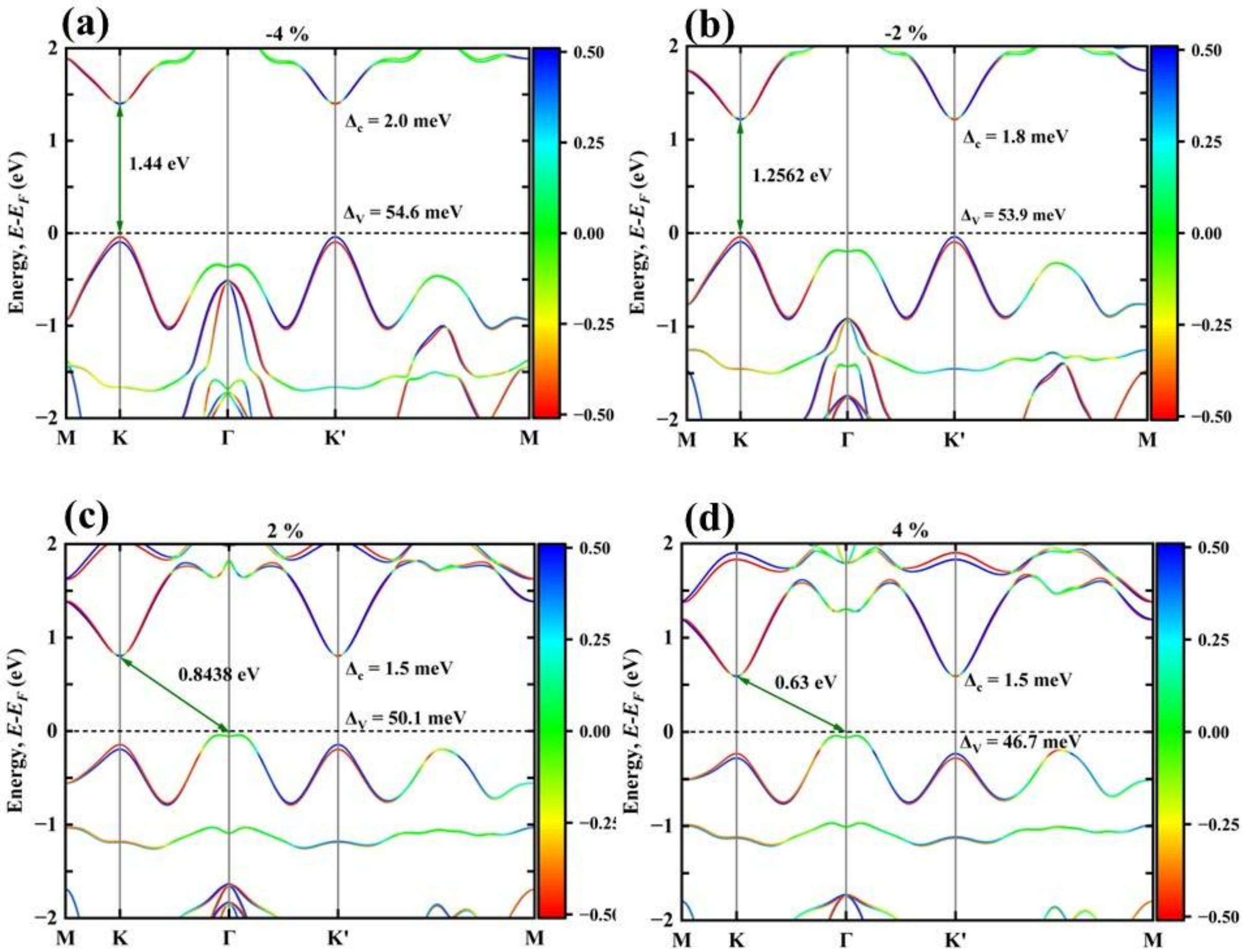

**Figure 10.** Spin-resolved band structures of monolayer $CrGeCN_4$ with SOC under biaxial strains of (a) −4%, (b) −2%, (c) +2%, and (d) +4%. The color scale represents the out-of-plane spin expectation value $\langle \mathbf{S_z} \rangle$, indicating strain-induced spin splitting and valley polarization.

In the absence of strain, $CrSiCN_4$ exhibits a direct band gap of 1.23 eV at the K valley, whereas $CrGeCN_4$ possesses a smaller direct gap of 1.09 eV. Compressive strain enhances the band separation, increasing the gaps to 1.59 eV and 1.44 eV at -4% strain for $CrSiCN_4$ and $CrGeCN_4$, respectively. In contrast, tensile strain progressively narrows the gaps, reaching 0.86 eV and 0.63 eV at +4% strain. This monotonic reduction originates from the strain-induced modification of orbital hybridization and crystal-field splitting, which alters the relative energetic positions of the CBM and VBM. Notably, tensile strain beyond +2% drives a direct-to-indirect band-gap transition in both systems, providing an effective route for strain-controlled optical and electronic functionalities.

The inclusion of SOC reveals pronounced spin splitting near the valley edges with opposite spin orientations at the time-reversal related K and K′ valleys, demonstrating robust valley-contrasting spin polarization. The conduction-band splitting remains relatively weak and nearly strain independent, with $\Delta_c$ values of 1.5-1.6 meV for $CrSiCN_4$ and 1.5-2.0 meV for $CrGeCN_4$. In contrast, the valence-band splitting reaches considerably larger values of 49.5-54.2 meV and 46.7-54.6 meV, respectively. The pronounced asymmetry between $\Delta_v$ and $\Delta_c$ indicates a stronger SOC contribution to the valence band states, while the nearly invariant splitting under strain suggests that the intrinsic spin-valley coupling is preserved against moderate lattice deformation. The comparable spin-splitting magnitudes in both compounds indicate that the Cr-based electronic framework plays the dominant role, while Ge substitution mainly provides additional chemical modulation through its larger atomic size and stronger atomic SOC contribution.

Beyond the modulation of electronic and spin characteristics, biaxial strain provides an effective pathway for engineering the carrier dynamics through the reconstruction of band-edge dispersion. The calculated electron and hole effective masses along the K-Γ and K-M directions (**Figure 11(c, d**)) reveal pronounced anisotropic transport characteristics in both $CrSiCN_4$ and $CrGeCN_4$. The electron effective mass exhibits a systematic reduction along the K-Γ direction under deformation, reflecting enhanced band curvature and increased electronic dispersion. For instance, in $CrSiCN_4$, the electron effective mass decreases from $1.51m_0$ at -4% strain to $0.87m_0$ at +4% strain, accompanied by a similar evolution in $CrGeCN_4$. In contrast, the hole effective masses display a more non-monotonic and direction dependent response, highlighting the distinct sensitivity of valence band states to lattice deformation. This contrasting behavior between electron and hole carriers demonstrates that strain does not merely tune the magnitude of the band gap but actively reshapes the momentum-dependent electronic landscape governing carrier transport. The observed anisotropic effective-mass evolution originates from strain-induced modifications of valley dispersion and band-edge curvature, providing an additional degree of freedom for directional carrier control. Therefore, strain engineering in $CrSiCN_4$ and $CrGeCN_4$ offers a versatile strategy to tailor electronic transport characteristics and complementing their robust spin-valley properties for multifunctional two-dimensional valleytronic applications.

Overall, these results establish strain as an efficient strategy for engineering the coupled electronic, spin, and transport properties of $CrSiCN_4$ and $CrGeCN_4$. The coexistence of a widely tunable band gap, robust valley-contrasting spin polarization with sizable valence-band splitting (~ 50 meV), and strain-dependent carrier anisotropy highlights these two-dimensional materials as promising platforms for strain-controlled valleytronic, and optoelectronic applications.

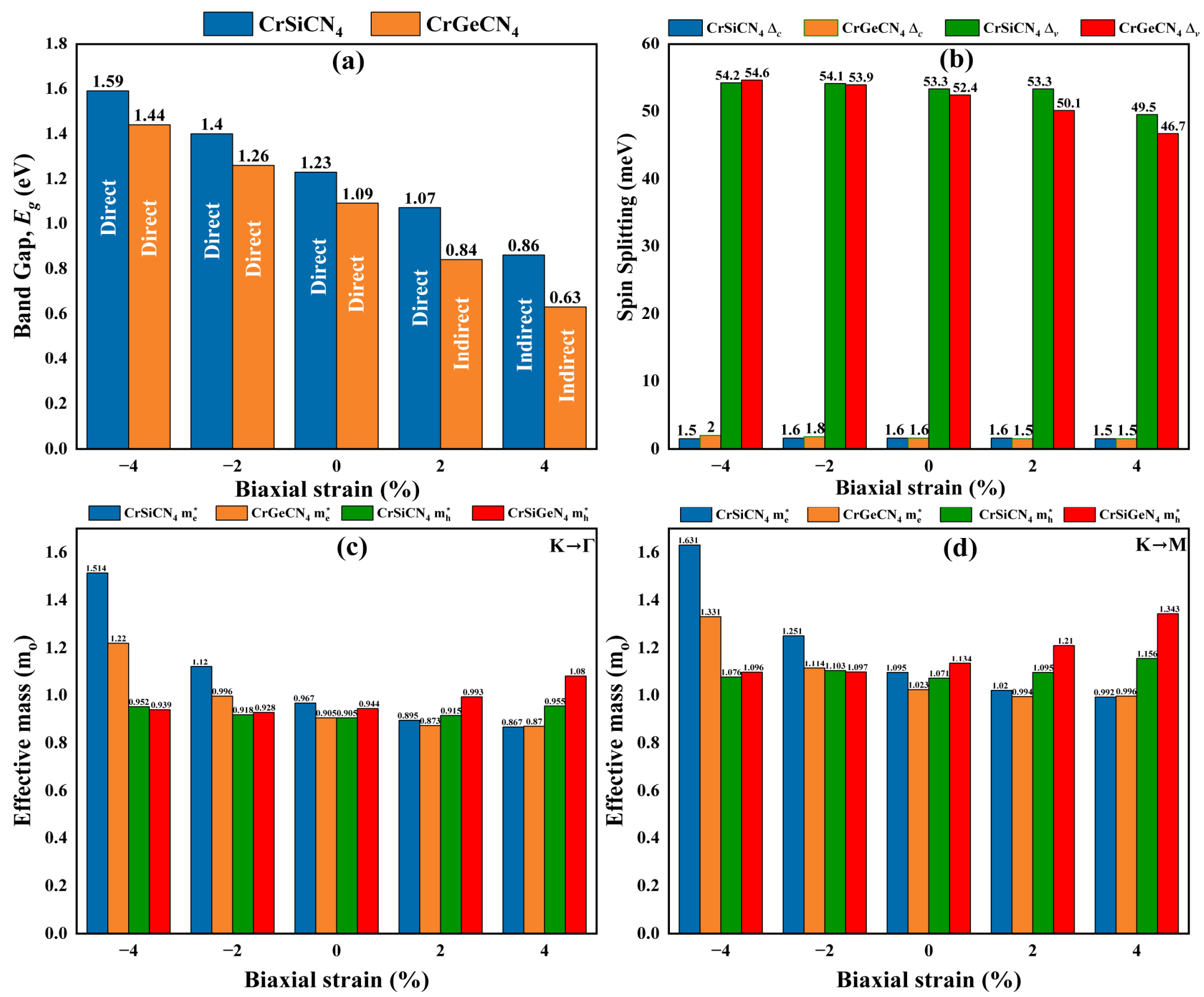

**Figure 11.** Effect of biaxial strain on the electronic and spin-dependent properties of monolayer $CrSiCN_4$ and $CrGeCN_4$. (a) Strain-induced modulation of the band gaps. (b) Variation of spin splitting at the conduction and valence band edges. (c, d) Evolution of carrier effective masses along the K-Γ and K-M directions, respectively, under different strain conditions.

## 3.8 Strain-dependent optical absorption

Mechanical strain is an established route for tuning the electronic and optical energy scales of two-dimensional semiconductors: tensile strain lowers the optical gap of $MoS_2$ [49], while first-principles calculations show that biaxial strain can strongly modify the band gaps of related $MX_2$ monolayers [50]. Strong visible-light absorption has also been reported for Janus MoSSe [51], PtSSe [52], and group-III chalcogenide monolayers [5]. **Figure 12** compares the calculated optical absorption spectra of $CrSiCN_4$ and $CrGeCN_4$ under biaxial strains from -4% to +4%. At equilibrium, $CrSiCN_4$ exhibits a pronounced low-energy peak at 2.92 eV with an absorption of 18.86%, followed by its dominant maximum at 9.66 eV (25.54%). The corresponding $CrGeCN_4$ features occur at lower energies, 2.64 eV (18.01%) and 9.16 eV (24.26%), respectively. Thus, relative to $CrSiCN_4$, $CrGeCN_4$ shows a consistent displacement of the main spectral features toward lower photon energy without a uniform increase in peak magnitude.

The lower-energy response is particularly sensitive to strain. The first pronounced peak of $CrSiCN_4$ shifts from 3.24 eV at -4% to 2.48 eV at +4%, whereas the corresponding feature in $CrGeCN_4$ moves from 2.84 to 2.24 eV. Within the shaded visible interval (~ 1.65-3.10 eV), the mean optical absorption increases from 7.92% to 11.20% for $CrSiCN_4$ and from 10.16% to 11.88% for $CrGeCN_4$ as the strain changes from -4% to +4%. Broad extension of light harvesting toward lower photon energies has been reported in other two-dimensional systems [**53**], while structural modification can also redshift prominent optical features [**54**]. In closely related Janus $XMSiY_2$ monolayers, tensile strain was reported to redshift the absorption onset, whereas compressive strain shifted it toward higher energies [**55**], in qualitative agreement with the strain-dependent optical response observed in the present $CrMCN_4$ monolayers.

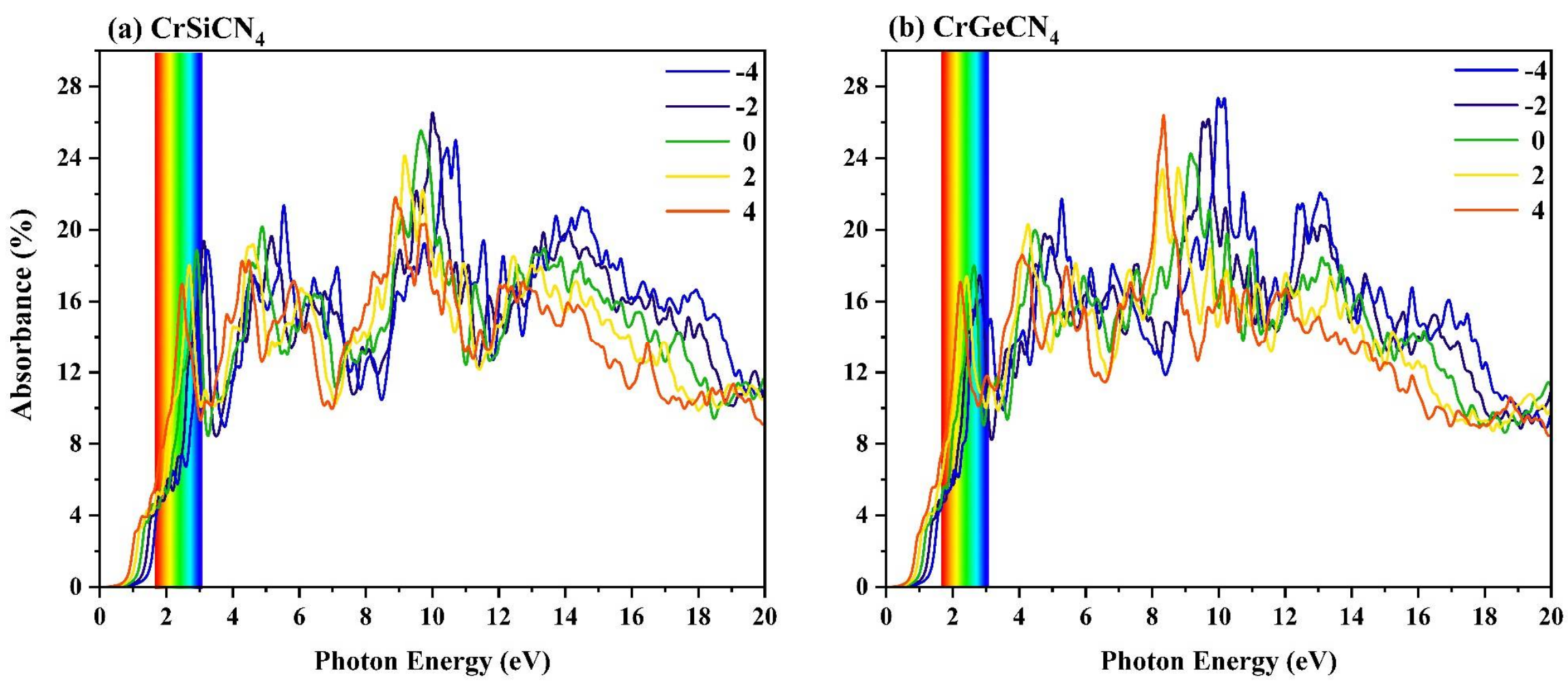

**Figure 12.** Strain-dependent optical absorption spectra of (a) $CrSiCN_4$ and (b) $CrGeCN_4$ monolayers under biaxial strains of -4%, -2%, 0%, +2%, and +4%. The shaded region denotes the visible-light range.

Strain redistributes the higher-energy part of the spectra rather than simply scaling the curves. The robust picture is therefore a strain-controlled migration of the dominant absorption features, accompanied by material-dependent intensity redistribution. The spectra for $CrGeCN_4$ remains generally shifted toward lower photon energy and shows the larger average response in the visible window, whereas that of $CrSiCN_4$ displays the slightly larger total shift of its dominant peak across the studied strain range. These trends indicate a tunable optical spectral window under moderate biaxial strain and may be relevant to strain-controlled optoelectronic design. The absorption spectra alone, however, do not establish specific orbital transitions, excitonic character, photovoltaic efficiency, photocurrent, or complete device performance.

# 4 Conclusions

Janus $CrSiCN_4$ and $CrGeCN_4$ monolayers combine spin-valley coupling with composition-dependent Berry curvature and complementary piezoelectric responses. HSE06 hybrid-functional calculations confirm that both unstrained monolayers are direct-gap semiconductors, with band gaps of 1.46 eV for $CrSiCN_4$ and 1.19 eV for $CrGeCN_4$. Cr-*d*/N-*p* hybridization dominates the frontier states, with orbital-dependent spin-orbit coupling producing valence-band spin splitting near 53 meV and conduction-band splitting near 1.6 meV. Opposite out-of-plane spin character and Berry curvature at K and K′ provide the basis for valley-dependent transverse transport while preserving time-reversal symmetry. The much stronger Berry-curvature hot spots in $CrGeCN_4$, despite comparable valley-edge spin splittings, point to a role for substitution-sensitive multiband coupling.

Biaxial strain from −4% to +4% narrows the PBE+SOC gaps from 1.59 to 0.86 eV in $CrSiCN_4$ and from 1.44 to 0.63 eV in $CrGeCN_4$. Tensile strain induces indirect-gap behavior while retaining K/K′ spin contrast, modifies carrier effective masses, and shifts absorption toward lower photon energies. Ge substitution increases the in-plane piezoelectric coefficient $d_{11}$ from 3.305 to 5.084 pm $V^{-1}$, whereas $CrSiCN_4$ exhibits the larger out-of-plane magnitude, $|d_{31}|$ = 0.082 pm $V^{-1}$, compared with 0.020 pm $V^{-1}$ for $CrGeCN_4$. The unstrained $CrSiCN_4$ band alignment remains insufficient for hydrogen evolution, making further band-edge engineering necessary for overall water splitting. These results provide a microscopic basis for selecting composition and strain to develop Cr-based Janus materials for valleytronic, optoelectronic, and electromechanical applications.

## Data availability

The data sets generated and/or analyzed in this study are available from the corresponding author on reasonable request.

## Declaration of interest

The authors declare no competing interests.

## CRediT authorship contribution statement

**Jubair Hossan Abir:** Conceptualization, Software, Methodology, Formal analysis, Data curation, Visualization, Writing-original draft, review & editing; **Md Rabi Us Sani Shapon:** Software, Formal analysis, Data curation, Visualization, Writing-draft, review & editing; **Sarker Sukanta Babu Pallab & Tanvir Khan:** Formal analysis, Visualization, Writing- draft; **Raihana Shams Islam:** Conceptualization, Supervision, Validation, Administration, Writing- Reviewing and Editing; **Saleh Hasan Naqib:** Conceptualization, Supervision, Validation, Administration, Writing- Reviewing and Editing.

## Appendix A

## Wannier interpolation

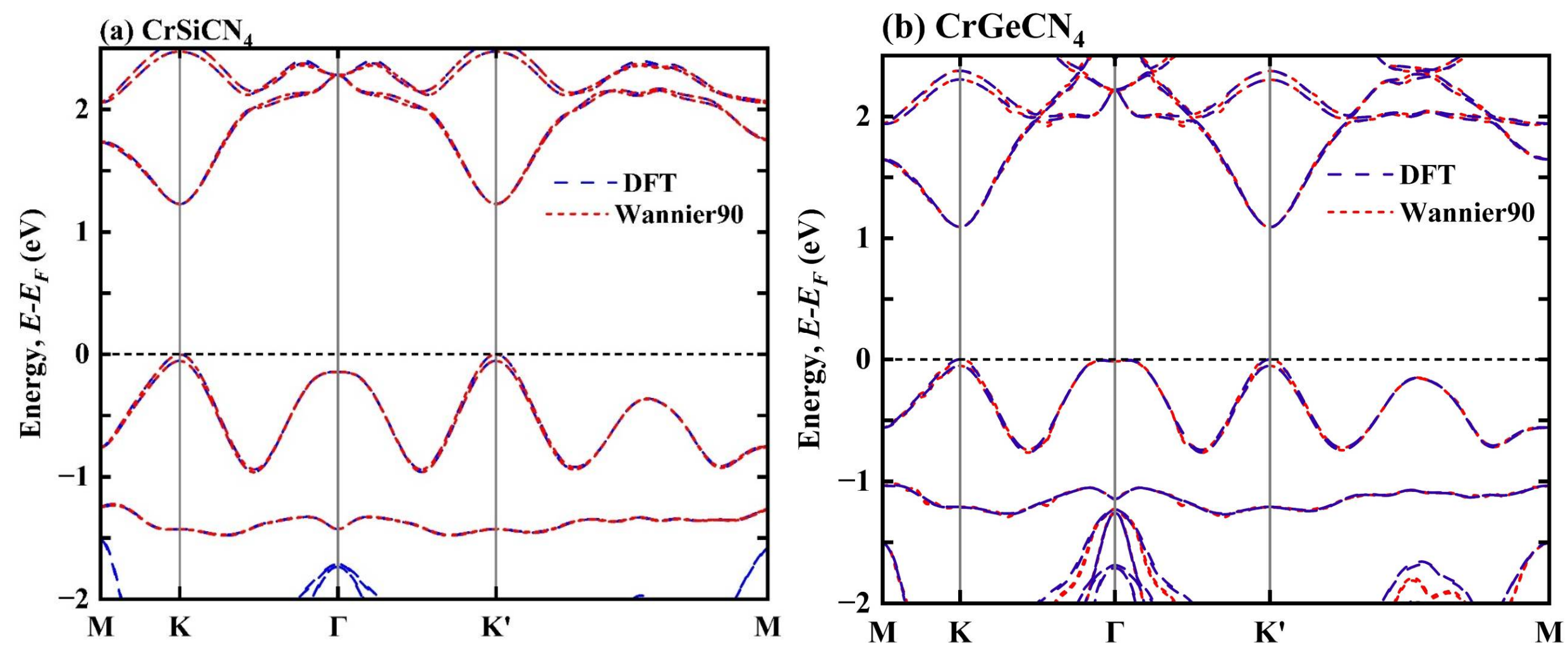

**Figure 13.** Figure shows a side-by-side comparison of electronic band structures derived from conventional DFT methods (dashed blue) and those reconstructed using maximally localized Wannier functions (dashed red)

## Spin-Polarized Electronic Structure without SOC

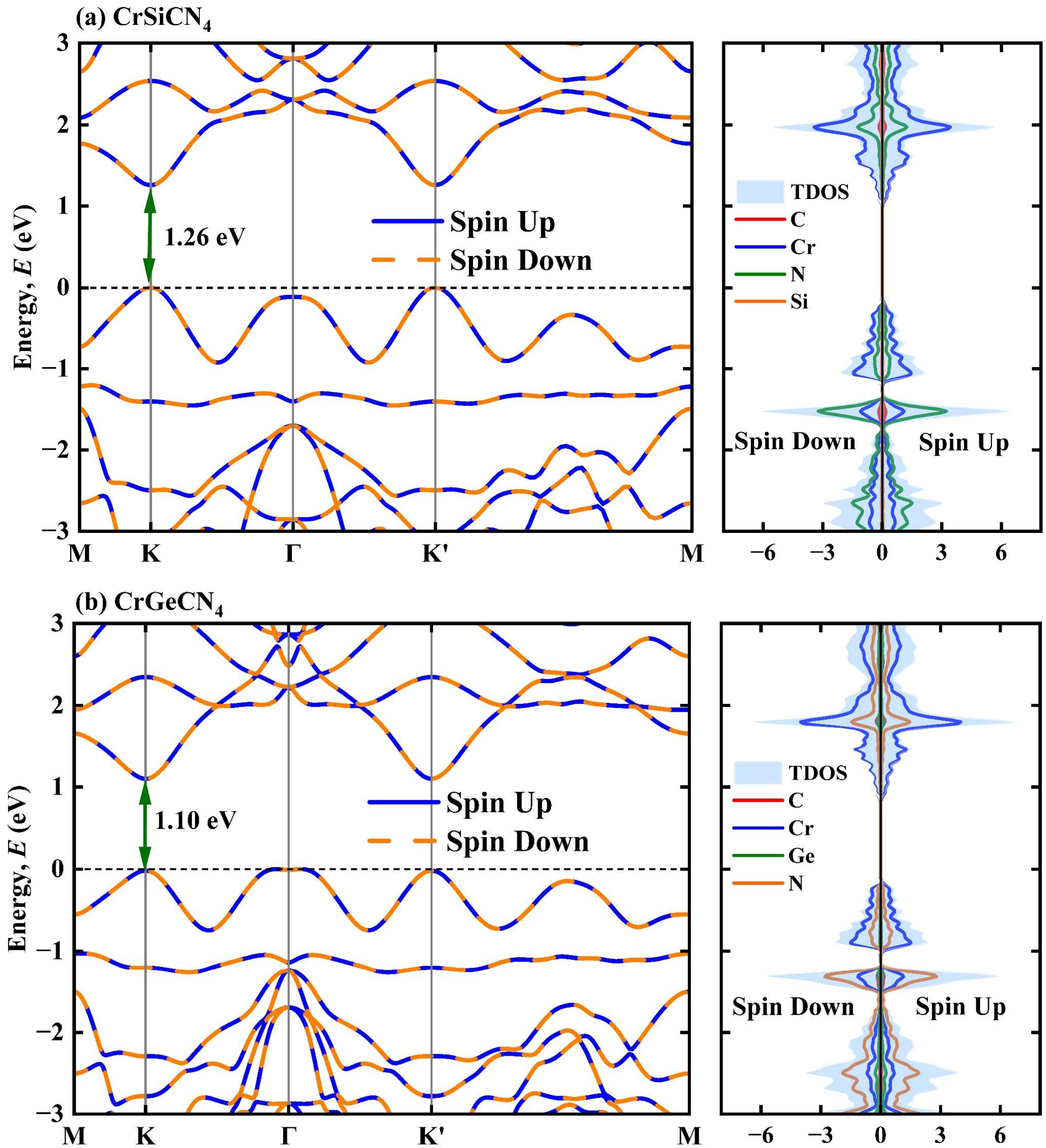

**Figure 14.** Spin-resolved band structure of monolayer (a) $CrSiCN_4$ and (b)$CrGeCN_4$. The right-side panel represents Spin-resolved projected density of states (PDOS). The Fermi level was set to 0 eV.